# Tackling problems, harvesting benefits – A systematic review of the regulatory debate around AI

by Anja Folberth[1,2], Jutta Jahnel[1], Jascha Bareis[1], Carsten Orwat[1], Christian Wadephul[1]

KIT SCIENTIFIC WORKING PAPERS 197

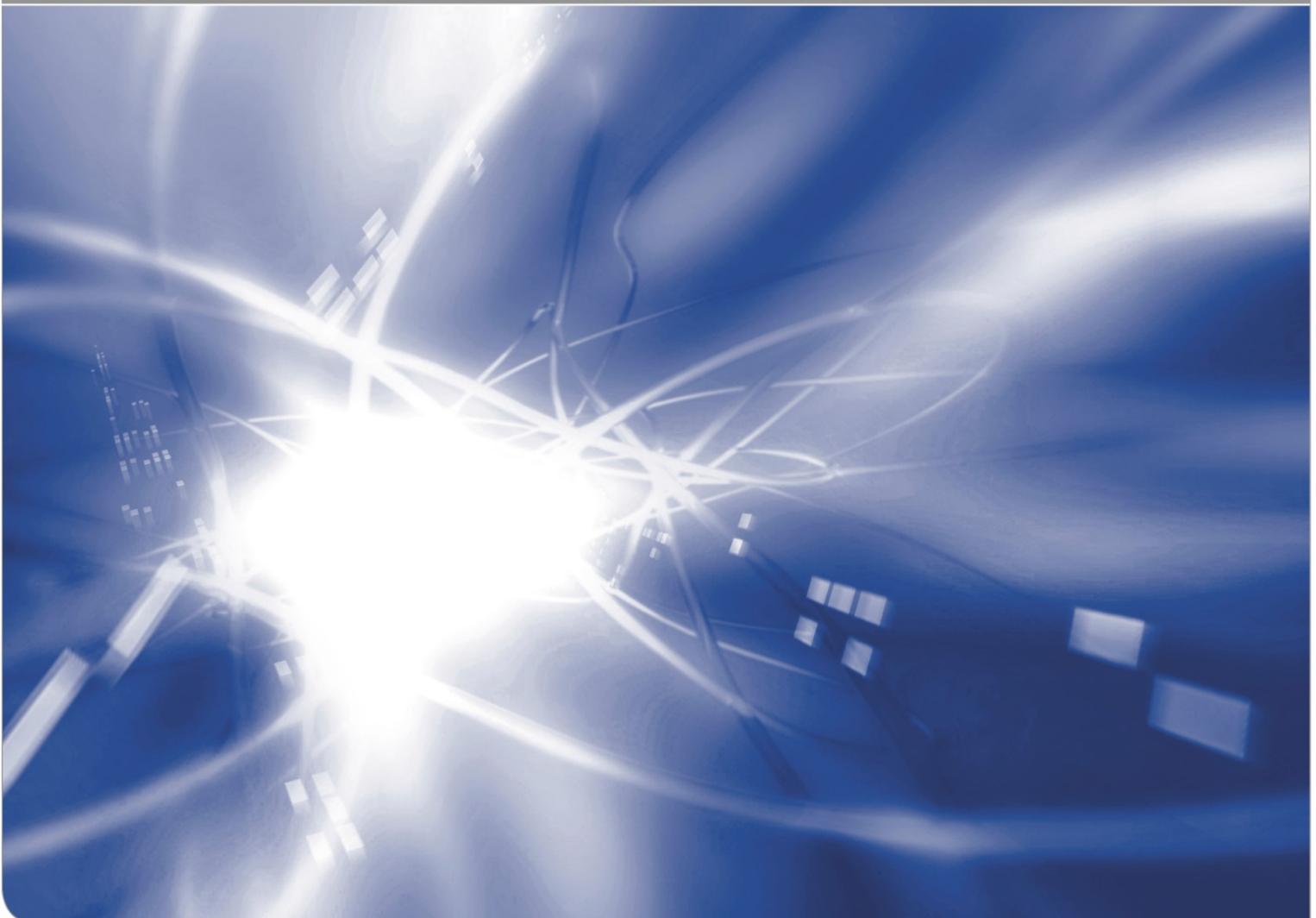

[1] Karlsruhe Institute of Technology, Institute for Technology Assessment and Systems Analysis (AF and CW until 2021)

[2] Institute for Political Science, University of Heidelberg

**Funding**

This work was supported by the German Federal Ministry of Education and Research (Grant number 01IS19020B, research project "Governance von und durch Algorithmen" (GOAL)). The funding is gratefully acknowledged.

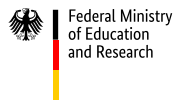

**Address**

Institute for Technology Assessment and Systems Analysis (ITAS)
P.O. Box 3640
76021 Karlsruhe
Germany
https://www.itas.kit.edu/

**Suggested citation**

Folberth, Anja; Jahnel, Jutta; Bareis, Jascha; Orwat, Carsten; Wadephul, Christian (2022): Tackling problems, harvesting benefits – A systematic review of the regulatory debate around AI; KIT Scientific Working Papers No. 197, Karlsruhe: Karlsruhe Institute of Technology (KIT), Institute for Technology Assessment and Systems Analysis (ITAS)

**Impressum**

Karlsruher Institut für Technologie (KIT)
www.kit.edu

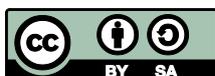

2022

ISSN: 2194-1629

# Tackling problems, harvesting benefits –
# A systematic review of the regulatory debate around AI

**Abstract**
How to integrate an emerging and all-pervasive technology such as AI into the structures and operations of our society is a question of contemporary politics, science and public debate. It has produced a considerable amount of international academic literature from different disciplines. This article analyzes the academic debate around the regulation of artificial intelligence (AI). The systematic review comprises a sample of 73 peer-reviewed journal articles published between January 1st, 2016, and December 31st, 2020.

  The analysis concentrates on societal risks and harms, questions of regulatory responsibility, and possible adequate policy frameworks, including risk-based and principle-based approaches. The main interests are proposed regulatory approaches and instruments. Various forms of interventions such as bans, approvals, standard-setting, and disclosure are presented. The assessments of the included papers indicate the complexity of the field, which shows its prematurity and the remaining lack of clarity. By presenting a structured analysis of the academic debate, we contribute both empirically and conceptually to a better understanding of the nexus of AI and regulation and the underlying normative decisions. A comparison of the scientific proposals with the proposed European AI regulation illustrates the specific approach of the regulation, its strengths and weaknesses.

# Table of contents

# List of tables

# List of figures

# 1 Introduction

Nowadays, Artificial Intelligence (AI) penetrates all spheres of life, ranging from product recommendation engines, speech and image recognition (Lauterbach, 2019: 241), to spam filtering, traffic planning, logistics management, and diagnosing diseases (Zuiderveen Borgesius, 2020: 1572), as well as the prediction of criminal behavior, and court decisions (Bloch-Wehba, 2020: 1267).

Publicly discussed incidents connected to discrimination, breach of confidentiality and financial loss, unravel how data processing activities can infringe upon basic rights and values. This confronts political stakeholders with ever more urgency to advance a regulatory framework. Recent European AI policy proposals explicitly aim at closing this regulatory gap (European Commission, 2021b). Further, the technological and economic potentials of AI, together with the multidimensional risks, have provoked a considerable amount of scholarly work in recent years. Nevertheless, Robles Carrillo lamented that there is still "relatively limited research on the overall panorama of the ethical or legal problems posed by AI" (Robles Carrillo, 2020: 2).

There is published work on systematizing proposals of ethical guidelines and principles (e.g., Jobin et al., 2019; Rudschies et al., 2021), but it does not focus on encompassing regulatory proposals. Furthermore, a recently published review analyzes existing regulatory approaches and models, including laws and governmental strategies for the development of an integrative process framework for the governance of AI (de Almeida et al., 2021). Its framework is based on a broad strategy embedding political and social actors and introducing a new regulatory agency. In this review, we aim to enlarge the discussion by focusing on concrete regulatory instruments. By conducting a systematic review of peer-reviewed journal articles, we map the academic debate and give an overview of proposed regulatory instruments.

We firstly point out to the debate of the term 'regulation' and present the European *Proposal for a Regulation on Artificial Intelligence (Artificial Intelligence Act, hereinafter: proposed AIA)* (European Commission 2021b) as an example of regulating AI (Section 2). Section 3 describes the methodological approach and the sample selection, including methodological limitations. We present the specific results of the analysis in Section 4 and follow up with an overarching discussion considering the proposed AIA (Section 5). Finally, we present our conclusions in Section 6.

# 2 Background – towards a working definition of regulation and a concept for regulating AI

## 2.1 Regulation

Broad scholarly interest and research reinforce the contestation and confusion around concept of regulation (Black, 2002: 11; Black and Murray, 2019: 9). Some authors even state that it might be impossible to come to a single definition of the term regulation (Baldwin, 1998: 2). However, analyzing 101 articles from different social science disciplines, Koop and Lodge conclude that there is an implicitly shared minimal understanding of regulation as "intentional intervention in the activities of a target population". This understanding includes direct and indirect regulation as well as governmental and self-regulation frameworks (Koop and Lodge, 2017: 104).

We follow such minimal consensus, as it encompasses a broad spectrum of policy measures from different regulatory fields targeting AI, such as governmental interventions (e.g., via bans or moratoria), and corporate self-regulatory frameworks, but also public participatory initiatives.

## 2.2 Risk-based regulation – a prominent concept for the regulation of AI

Currently, one prominent approach we see when bringing together regulation and AI is the risk-based approach that focuses on handling the adverse potential impacts of technology applications. Not only parts of the academic debate refer to risk-based approaches (e.g. German Data Ethics Commission, 2019; Krafft et al., 2020), but also policy-makers such as the European Commission (2021b) with the proposed AIA. The risk-based approach provides a decision-making framework and procedures to prioritize regulatory activities and the deployment of resources according to the risk levels ascribed to different regulatory objects (Black, 2010).

Within the proposed AIA, the risk-based approach attributes different levels of risks of AI applications to corresponding regulatory measures. AI systems with 'unacceptable' risk levels, i.e., certain manipulative systems, social scoring systems, or some remote biometric identification systems, would be prohibited (Art. 5 proposed AIA). Providers of 'high risk' AI systems would have to establish and operate various regulatory means. This includes a quality management system with a documented risk management system, data management practice, and obligations on accuracy, robustness and cybersecurity. Furthermore, technical documentation and logging for traceability, provision of information to users, human oversight, conformity assessment as a form of self-assessment by the providers, a CE marking of conformity, the registration of the systems and a post-market monitoring are proposed. A third-party conformity assessment by notified bodies is stipulated for some systems for biometric identification and categorization of natural persons as well as for AI systems that are safety components of products covered by certain EU safety legislations that usually require conformity assessments by notified bodies (Chapter 2 and 3 proposed AIA). For AI applications with a risk level deemed as 'limited' risk, such as AI systems that interact with natural persons, emotion recognition systems, biometric categorization systems, or 'deep fake' systems, there would only be a duty to inform about the operation of the system or to disclose that the content is manipulated (Art. 52 proposed AIA). Additionally, the voluntary application of codes of conduct is suggested for AI systems with 'minimal' risks (Art. 69 proposed AIA). A further regulatory measure is continuous market surveillance and control, both as post-market monitoring by providers (Art. 61 proposed AIA) and by market surveillance authorities (MSAs). For MSAs, the proposal provides certain means of access to data and the source code as well as means to evaluate AI systems, to require appropriate corrective measures, to prohibit or restrict AI systems being made available on the market, and to withdraw or recall them (Art. 63-68 proposed AIA).

## 3 Methodology

### 3.1 Collection and selection of papers

To collect our sample, we scanned five databases: Web of Science, Scopus, Worldwide Political Science Abstracts, PAIS Index, and Sociological Abstracts (see Figure 1). We designed the following search terms to include publications aiming at the intersection of regulation and AI (Figure 2; for detailed search operators see supplementary Appendix): *"regulat\*", "governance", "policy mak\*", "politic\*", "legislat\*", and "AI", "artificial intelligen\*", "algorithm\*", "ADM", "automated decision", "machine learning", "ML", "deep learning", "intelligent system", "autonomous system", "expert system".*

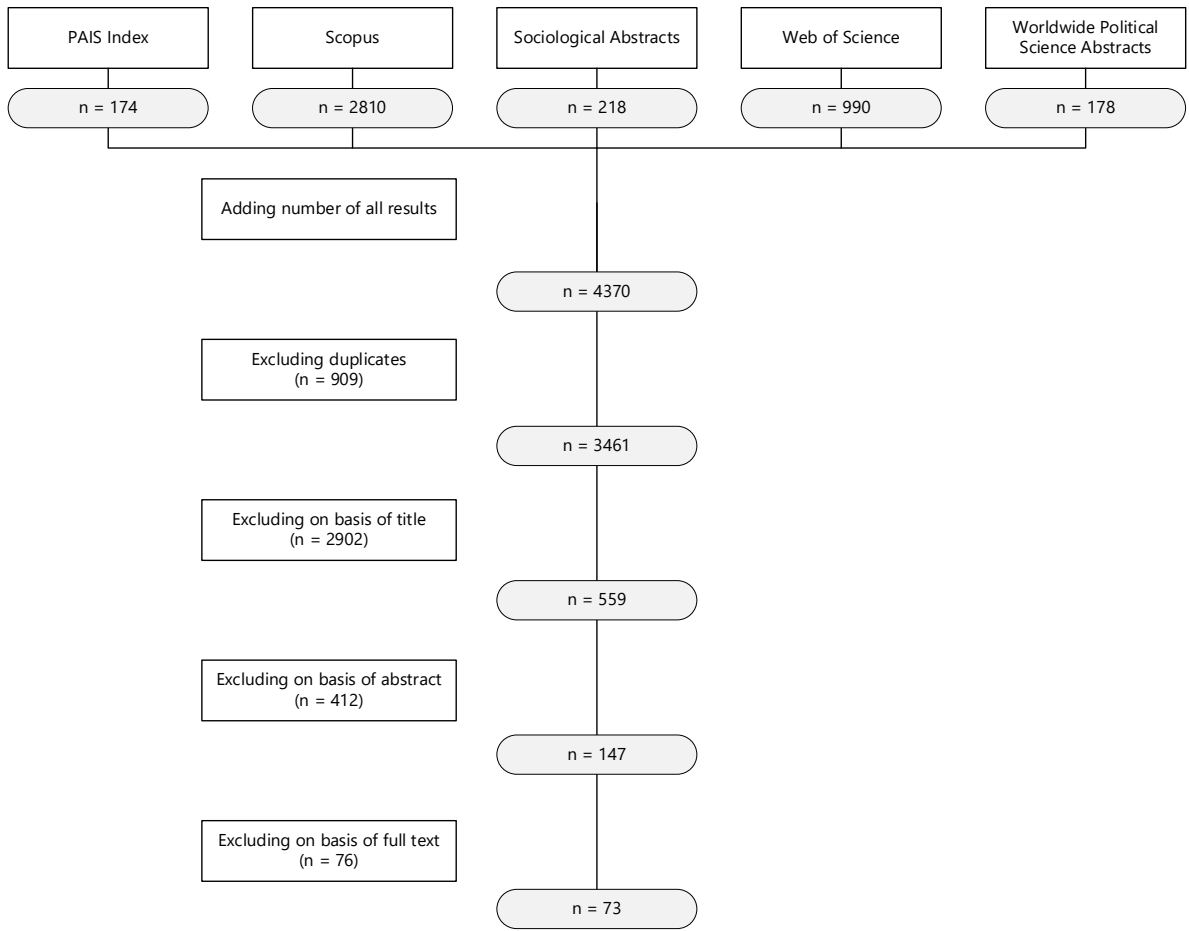

Figure 1: Article selection process

Further inclusion criteria were: (1) English language; (2) peer-reviewed journal articles; (3) papers published between January 1st, 2016, and December 31st, 2020. To narrow down the outcome to relevant results we excluded some subjects (exact selection depending on database design, see Appendix). Subsequently, we sorted the remaining results manually on the basis of title, abstract, and full-text reading (Figure 1). The sample selection process, as well as the other methodological features of the systematic review, follow the PRISMA-P 2015 checklist (Moher et al., 2015: 5).

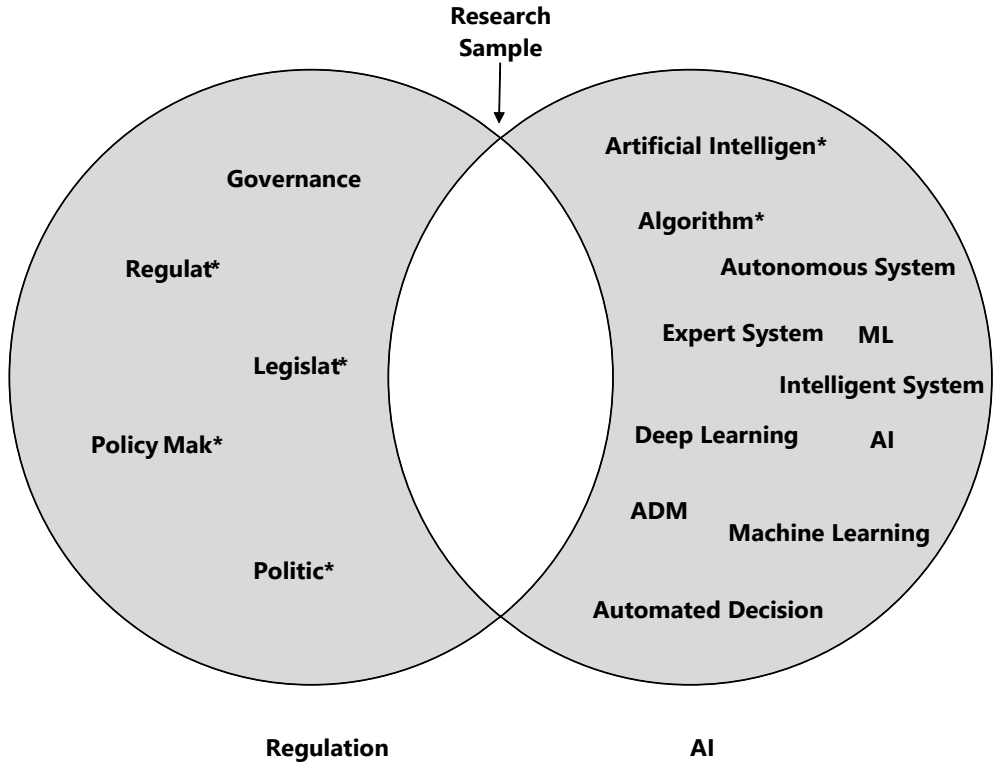

Figure 2: Keywords for search operators

We focused on broad, horizontal concepts of regulating AI as a reaction to the currently limited research on an overall picture in this area, that results in "a sectorial and fragmented perspective prevail[ing] over the integral and holistic overview" (Robles Carrillo, 2020: 2). Therefore, we did not include articles that were particularly concerned with regulating specific sectors (e.g., health, education, work, etc.) or salient issues such as autonomous driving, lethal autonomous weapon systems, deep fakes, or robotics. These issues are widely discussed and need targeted systematic reviews.

Out of 4.370 search results, we excluded 909 duplicates in the first step. In the second step, the first author excluded articles based on titles that did not refer to a political dimension of regulation (e.g., chemical processes, medical research, engineering issues, mathematical models) and accidental matches concerned with issues that do not match the topic of the review.

In the third step, the first author excluded articles that did not give a broad view on regulating AI out of the remaining 559 articles by reading the abstracts. Articles that did not precisely fit the exclusion criteria were included in the next selection step. In the final step, four of the authors excluded another 76 of 147 articles based on full-text reading. Only articles which proposed regulatory instruments were included in the final sample of 73 articles.

We analyzed our final sample with MAXQDA, a software specialized for qualitative research. Our procedure followed deductive reasoning as we implemented a preliminary code structure before starting to code. During this procedure we developed the detailed structure inductively by adjusting and adding further important dimensions. Three of the authors coded the same seven articles separately to compare and combine prioritization and to develop the first structure together. Afterwards, four authors coded articles openly.

## 3.2 Methodological limitations

We chose a two-eyes-principle instead of the recommended four-eyes-principle in selecting relevant articles based on title and abstracts, as there were more than 4.000 results. To prevent subjective bias and missing relevant articles, we always included ambiguous cases in the next selection step. National discourses may have been missed, as we only analyzed articles published in English due to language knowledge of the author team. Additionally, we focused on peer-reviewed articles and did not include book sections or reports and grey literature that might add even more details and ideas to the discourse.

## 4 Results

In this section, we present our results starting with some general remarks on our final sample. A first finding points to the increasing relevance of regulating AI during our research sample period (see Figure 3). We started the review with articles published in 2016, a year where AI related affairs were trending in the news. The AI system AlphaGo defeated Fan Hui, the established champion of the complex game Go (Nindler, 2019: 12). Joshua Brown's death in an accident caused by his semi-autonomous Tesla vehicle fueled the *public discussion* on safety of autonomous driving (Abdul Manap and Abdullah, 2020: 187). In the judiciary field, Angwin et al. (2016) showed that the COMPAS system applied in US jurisdiction discriminates against black people. Besides to the public debate, *politics* started to address AI systems. While the White House published a first report on AI, the negotiations on the General Data Protection Regulation (GDPR) in the European Union entered the final phase. The final version published in April 2016 allows automated decision-making and profiling only under specific conditions.

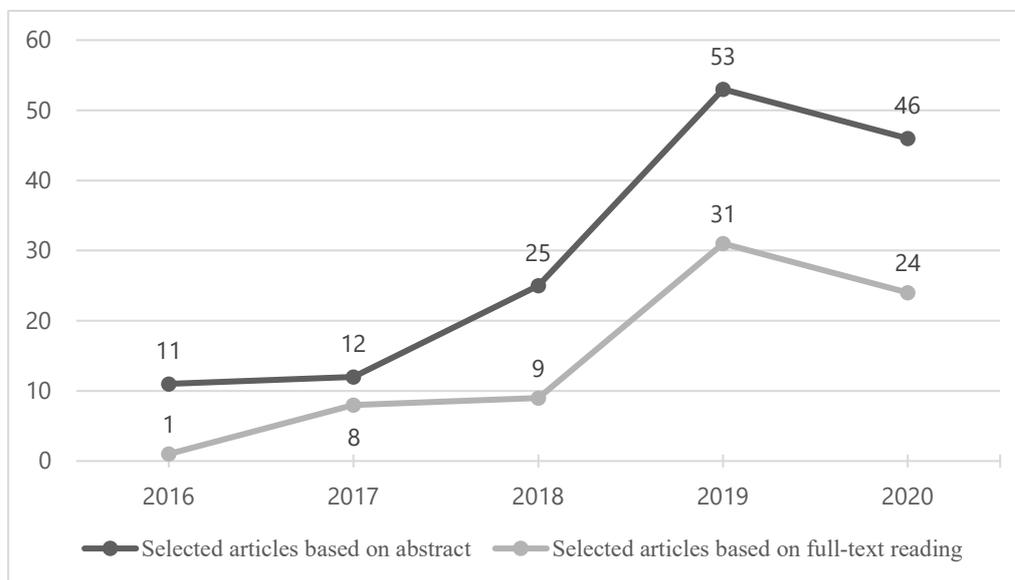

Figure 3: Number of articles related to the topic of the review

This topical trend around AI set our starting point, and we found 11 articles of interest in the *academic debate* of 2016, but most of them were concerned with particular problems. In addition, we found widespread mapping of problematizations that did not yet include regulative proposals (Zarsky, 2016; Ziewitz, 2016). Only one article from 2016 was included in the final sample (Hirsch, 2016), followed by an increasing number of articles the following years (Figure 3). We conclude that 2016 is a starting point of a bigger regulatory debate.

### 4.1 Regulation object – in need of specification of AI

Our findings show that there is a need to define the regulation object, as a third of the papers problematize the terminology around AI as fuzzy. Most of the articles relate to automated decision-making (ADM) or machine learning (ML) systems explicitly, some do not specify AI at all. Only a few papers attempt a clear-cut definition of terms. Some definitions derive from reports or policy documents, like the OECD Recommendations (Gacutan and Selvadurai, 2020: 208), the EU High Level Expert Group on Artificial Intelligence (Bannister and Connolly, 2020: 471-472; Smuha, 2020: 1), or the GDPR (Goodman and Flaxman, 2017: 1; Brkan, 2019: 93). Others refer to short dictionary definitions (Brand, 2020: 117-118; Butcher and Beridze, 2019: 88, Cerrillo i Martínez, 2019: 16) or make their own attempts to define AI (Abdul Manap and Abdullah, 2020: 191) and related terms (Bannister and Connolly, 2020: 471; Mann and Matzner, 2019: 1).

In the regulation debate, further scrutiny in differentiating between a general AI definition and the regulation object is pivotal. Targeting the regulation object often narrows down to classifications of AI systems to determine what kind of systems and applications need to be regulated. Robles Carrillo states that there is not enough attention to the difference in typologies of AI systems, which "cannot be explained or treated equally" (Robles Carrillo, 2020: 9). The review results show very few articles that try to circumscribe a regulation object. Bannister and Connolly demand a regulation for "decision-making algorithms which have the potential to materially affect citizens, business or other organisations on a continuing basis" which excludes "for example, decision support systems or models built for one-off decisions" (Bannister and Connolly, 2020: 487). Brkan distinguishes between high impact ADM systems: "when a decision is binding for individuals and affects their rights", e.g. concerning credit, tax return or employment, and low impact: "If the automated decision-making does not have any binding effect on data subjects and does not deprive them of their legitimate rights" (Brkan, 2019: 93-94).

### 4.2 Reasons for regulatory action – problems addressed

The results show a considerable number of problems and risks, which are described and characterized as giving reasons for possible levels of regulatory action. There is not one article that failed to address various types of problems and challenges which the use of AI or ADM entails. Even innovation-oriented articles which warned that "governmental or even self-imposed controls […] limit or slow down development of AI" (Etzioni and Etzioni, 2017: 33) and instead demanded the regulation of AI "by a policy regime of permissionless innovation" (O'Sullivan and Thierer, 2018: 33) express concerns. The most prominent concerns in the academic discourse are the infringement of rights, such as *illegal discrimination* or *privacy concerns*, and technical issues such as the *complexity* or *inaccuracy* of automated decisions. Especially the problem of *opacity*, also known as the "black box problem", is widely discussed. This problem arises when an algorithmic system makes decisions which are extremely difficult to explain to an average person. Furthermore, authors are concerned when it comes to the question of *accountability for decisions*, and the *necessity of human control* in supervising ADM systems. In addition, the degree of social agreement, or disagreement, that characterizes the *normative context under examination* are attributed as challenges for policymakers (e.g., Black and Murray, 2019: 11; Coglianese and Lehr, 2019: 8; Lauterbach, 2019: 238; Calo, 2017: 415; Gasser and Almeida, 2017: 59). An extensive body of research has also examined the impact of AI on the *transformation of employment*, and the potential influence on vulnerable groups by *disrupting political discourses* with implications for democracy, identity, and culture (e.g., Wang and Siau, 2019: 68; Bellanova, 2017: 333; Dignam, 2020: 48; Floridi et al., 2018: 692). The literature indicates that AI applications will tremendously transform human life (Nindler, 2019: 9). These concerns could be attributed to different kinds of possible risks as issues for regulation: more traditional risks of *physical or financial harm*, and specific new kinds of *psychological or broader social risk*, such as the general *transformation of the societal structure.* The various risks depend on the application context and are partly overlapping; especially the "black box issue" could be characterized as a systemic risk (Black and Murray, 2019: 7). Related to this, the analysis also shows that risks unfold at different target levels, as they are not limited to individuals but also relate to groups, systems (economic, political or science) and/or to the society in general (e.g., Lauterbach, 2019: 244; Bellanova, 2017: 333; Gasser and Almeida, 2017: 59; Calo, 2017: 431). Nindler even stated that AI in an advanced state (so-called *strong AI*) "could also pose an existential risk to humankind" (Nindler, 2019: 13).

The discussion shows a context-dependent and fragmented puzzle, rather than a clear-cut regulation trajectory with definite starting points and detailed strategies on how to handle risks and problems related to AI. However, despite the diversity of professional backgrounds and argumentation perspectives of the authors, there seems to be no disagreement that AI needs to be regulated.

### 4.3   Regulatory approaches and responsibilities

Most of the reviewed papers state that the appropriate form of regulation will vary depending on the societal deployment of AI. The suggested regulatory approaches mainly address the aforementioned problems and risks with a risk-based regulation, especially in the case of traditional risks (e.g., physical or financial). Some authors propose a more precautionary-oriented approach for cases of high risks and harms (Black and Murray, 2019: 13; Clarke, 2019: 406; Mannes, 2020: 65).

Regarding the normative justification of a regulation, questions arise concerning the adequate consideration of specific new types of risks posed by AI, e.g., mental or social risks through discrimination or the violation of human dignity. Some papers propose additional horizontal ethical rules and core principles based on common values and good governance, such as *privacy, transparency, accountability, autonomy, prevention of harm, fairness* and *justice* (Brkan, 2019: 93; Floridi et al., 2018: 696; Allen and Masters, 2020: 596; Cerrillo i Martínez, 2019: 15). According to Kozuka these principles should form the basic framework for the governance of AI (Kozuka, 2019: 328), which "are not a policy instrument or a legally binding regulation; but they appear to be [...] a kind of state-induced self-regulation" (Kozuka, 2019: 321).

Others propose that existing individual rights to get knowledge about automated decisions and options to redress (e.g., in the GDPR) should be complemented by collaborative governance (Kaminski, 2019: 1578), or through a "deontological approach to information privacy protection" (Gacutan and Selvadurai, 2020: 212). In particular, international human rights and fundamental rights are explicitly referred to (Black and Murray, 2019: 10; Feldstein, 2019: 50; Smuha, 2020: 595; Donahoe and Metzger MacDuffee, 2019: 116). For example, Brkan underlines the demand for a high level of transparency of data processing to achieve compatibility with the EU Charter of Fundamental Rights, in particular that "the data subject should have the right to know the reasons behind the decision" (Brkan, 2019: 120).

With regard to governance approaches, some authors explicitly recommend hybrid governance frameworks combining legal and non-legal norms, including additional regulatory players besides the governmental regulator (discussed with the terms "corporate governance" (Kozuka, 2019: 329; Felzmann et al., 2019: 10), "collaborative governance" (Kaminski, 2019: 1559), "co-regulatory framework", where "the detailed obligations are developed through a consultative process among advocates for stakeholders" (Clarke, 2019: 398; Carter, 2020: 3), or "multistakeholder governance" (Madhavan et al., 2020: 106; Munoko et al., 2020: 226; Brand, 2020: 122)). Hybrid governance structures are located on a spectrum between traditional governmental command-and-control regulation and voluntary private structuring and self-regulation, where law, soft law, and related policy mechanisms are deployed together (Kozuka, 2019: 316). It is argued that these governance structures should be enforced on a global level "to bridge gaps across different national legal systems" (Gasser and Almeida, 2017: 58).

This broad regulatory framework is specified by some authors: "Accountability for Reasonableness Framework" (Wong, 2019), "Polycentric Systems Framework" (Black and Murray, 2019: 14), framework to establish "Social Governance" (Lauterbach, 2019: 259) or a modular "Layered Governance Model" with social and legal, ethical and technical foundations (Gasser and Almeida, 2017: 60). Pure self-regulation is evaluated as ineffective by most of the reviewed papers, while some propose pure governmental regulation through existing hard law, or by creating new legislations or governmental oversight (Etzioni, 2018: 32; Kim, 2020: 931; Abdul Manap and Abdullah, 2020: 189; Abiteboul and Stoyanovich, 2019: 8). The formation of one or more special regulatory or advisory agencies on the national level which bring together "diverse stakeholders—from the open source community to commercial firms, to customers, to potential victims" (Tutt, 2017: 109) is considered to be a promising measure to allocate responsible actors and is proposed by many different authors (Abdul Manap and Abdullah, 2020: 191; Bannister and Connolly, 2020: 487; Butcher and Beridze, 2019: 88; Calo, 2017: 429; Cerrillo i Martínez, 2019: 22; Dignam, 2020: 48; Floridi et al., 2018: 403; Smith and Desrochers, 2020: 567; Treleaven et al., 2019: 39; Tutt, 2017: 90; Wirtz et al., 2020: 825). For instance, this special agency could be responsible for submitting AI products for testing and licensing (Dignam, 2020: 48). Other authors demand a new global or EU oversight body (Butcher and Beridze, 2019: 95; Cihon et al., 2020: 546; Donahoe and Metzger MacDuffee, 2019: 116; Gasser and Almeida, 2017: 61; Haenlein and Kaplan, 2019: 8; Nindler, 2019: 129; Robles Carrillo, 2020: 13; Floridi et al., 2018: 703), while Mannes (2020: 66) states that the expertise of already existing agencies should be increased.

## 4.4 Regulation instruments

Our key interest is to look at concrete regulatory instruments proposed due to the lack of adequate provisions or ineffective and unclear existing legislations.

Similarly to Favaretto et al. (2019: 15-16) in their systematic review on big data and discrimination, we identified some *technical solutions* suggested. Thirteen papers demanded an improvement of either the algorithm or the training data, and ten suggested the implementation of software to control the AI system's output, following the logic that "the first step to remove discrimination, according to this line of reasoning, is to increase the algorithm's accuracy" (Cofone, 2019: 1415). Most of the authors see technical improvement as one additional tool to prevent harm. Only very few papers, like Etzioni and Etzioni (2017), focus on finding technical solutions. Many others state that the challenges of AI cannot predominantly be solved with technical tools (Kaminski, 2019: 1584; Gasser and Almeida, 2017: 60; Floridi et al., 2018; Black and Murray, 2019; Kroll et al., 2017: 682). Butcher underlines, for example, to "ensure that effective governance is achieved, a mixture of technical tools and consensus-driven standards is required" (Butcher and Beridze, 2019: 96). The call for technical solutions is mostly recognized as a specific intervention in combination with *standard-setting*, *certification* or other instruments focusing on the measurement, evaluation and control of the *accuracy of outcomes*.

In the following subchapters, we clustered the variety of proposed regulatory instruments regarding different purposes.

### 4.4.1 Bans, approvals and standard-setting – regulatory actions for safety and accountability

Bans and approvals are among the most restrictive pre-market measures, also called "ex ante regulation". Eighteen articles suggest bans of technology applications, either explicitly or implicitly, if their implementation is not deemed appropriate. Some papers suggest bans of applications as a general tool (Krafft et al., 2020: 14; Lauterbach, 2019: 244; Mann and Matzner, 2019: 7; Smith and Desrochers, 2020: 572; Strandburg, 2019: 1881; Treleaven et al., 2019: 33; Turner Lee, 2018: 7). Others take bans of concrete technology applications into account, like biometric recognition systems (Bannister and Connolly, 2020: 474), "fully autonomous passenger flights" (Clarke, 2019: 406), autonomous weapon systems (Calo, 2017: 415; Etzioni, 2018: 31; Floridi et al., 2018: 698; Shin et al., 2019: 376), or AI systems that could be confused with humans due to their design, such as chatbots or voice assistance systems (Hoffmann and Hahn, 2019: 642).

Furthermore, we see proposals of pre-market approval that imply standards and requirements for technologies or applications to be adopted by developers and providers of AI systems. Only those systems which meet these provisions would then be permitted on the market. Some authors suggest pre-market approval without specifying the regulatory responsibility (Black and Murray, 2019: 13; Dignam, 2020: 39), while others directly connect pre-market approval to the duties of a special agency (Bannister and Connolly, 2020: 487; Smith and Desrochers, 2020: 571; Tutt, 2017: 111). Tutt (2017: 111) describes pre-market approval as "the most aggressive" position of a regulatory agency. Another suggestion is to combine pre-market approval with real-life laboratory testing of AI applications (Abdul Manap and Abdullah, 2020: 192; Floridi et al., 2018: 704; Lauterbach, 2019: 260).

One of the most prominent suggestions of further pre-market tools is *standard setting* for safe operation of ADM systems. In analyzing the details, we saw three main dimensions of standard setting: (1) design, (2) liability, and (3) performance. Design standards include debates around fairness, privacy, and responsibility by design. While design and performance standards aim at the accuracy of outcomes, liability standards regulate who is responsible for the prevention of possible harms. The proposals of liability standards are not discussed extensively in the context of AI regulation. There is no consensus or clear demand on who should be the responsible entity during the entire AI life cycle, reaching from the development stage to market placement and private usage, including developers, providers, natural, or legal persons. The papers only generally state that there is a need for a clear liability (Bannister and Connolly, 2020: 486; Brand, 2020: 121; Carter, 2020: 640; Hoffmann and Hahn, 2019: 636; Kaminski, 2019: 1573; Kozuka, 2019: 325; Treleaven et al., 2019: 39; Tutt, 2017: 109; Kim, 2020: 874). Mannes (2020: 65) differentiates between mandatory and voluntary standards, which have to be evaluated by an appropriate agency. Floridi et al. (2018: 702-703) put forward the idea of an insurance system to use market mechanisms and strengthen public trust in AI applications. Clarke expects "much more careful risk assessment and risk management" when "applying strict liability" (Clarke, 2019: 401). Rahwan argues in the same vein when expecting humans to avoid mistakes if they know they would bear the consequences. Therefore, he demands a human in the loop to "provide an *accountable entity* in case the system misbehaves" (Rahwan, 2018: 7).

Among those who suggest liability standards, some also demand easy access for complaints by affected individuals to challenge or dispute automated decisions (Allen and Masters, 2020: 597; Bannister and Connolly, 2020: 487; Carter, 2020: 6; Clarke, 2019: 406; Edwards and Veale, 2018: 13; Floridi et al., 2018: 702; Hoffmann and Hahn, 2019: 637; Smith and Desrochers, 2020: 571; Wong, 2019: 234). While setting standards on liability is a restrictive ex ante proposal, we also found suggestions to establish ex post liability regimes (Black and Murray, 2019: 13; Mannes, 2020: 65; Buiten, 2019: 56) through monitoring, controls, and auditing of permitted applications. AI systems would be allowed to enter the marketplace, with regulatory actions only needed in cases of failures.

Some papers also suggest certification to demonstrate compliance with existing regulations or standards. Systems that show the highest prospects for "catastrophic and deadly failures, such as aircraft" should undergo rigorous certification before real-world employment is permitted (Mannes, 2020: 65). In general, algorithms could be certified as software objects, by technology-based standards, and/or by performance-based standards. Another possibility is a certification of the legal person or the whole process which applies the system to make decisions. The advantage of certification standards lies in the sector-specific basis, which is already common in other areas, such as environmental sustainability standards (Edwards and Veale, 2018: 10). However, the value of the certificate as a guarantee of trustworthiness remains questionable in the domain of self-regulation, which is based on economic self-interests (Edwards and Veale, 2018: 10). Therefore, Kaminski (2019: 1568) only sees a possible effective compliance culture by structuring industry self-assessment through independent compliance officers and internal reports, or by involving independent oversight over a company's behavior, such as third-party auditors or civil society assessments.

Another part of the discussion about standard setting is the inclusion of ethics in developing and providing AI technologies. Many of the included papers refer to *ethics guidelines*, and some suggest implementing *ethical codes of conduct*, *ethics by design*, or *normative standards*. Many suggestions aim at self-regulation by implementing ethical principles, but there are also demands to impose obligations like *ethical training for software engineers* (Donahoe and Metzger MacDuffee, 2019: 124; Orr and Davis, 2020: 722; Truby, 2020: 955). However, others evaluate this kind of soft regulation as ineffective (Black and Murray, 2019: 8; Hoffmann and Hahn, 2019: 636; Shin et al., 2019: 376). Calo even warns that the "AI policy community must maintain a healthy dose of skepticism toward "ethical codes of conduct" developed by industry" (Calo, 2017: 424).

*4.4.2 Disclosure, labeling and information – regulatory actions for transparency*

Many included papers react to the problem of opacity and demand *different levels of disclosure*. Some consider a *hard revelation of technical details*, such as training data and models as pivotal (e.g., Buiten, 2019: 54; Hoffmann and Hahn, 2019: 637; Ienca, 2019: 275; Johnson, 2019: 1243; Krafft et al., 2020: 10; Tutt, 2017: 111). Others demand only a *soft revelation*, which means an explanation of the function of the algorithm that does not have to reveal code or training data details (e.g., Cofone, 2019: 1439; Coglianese and Lehr, 2019: 39; Strandburg, 2019: 1882; Zerilli et al., 2019: 676). In some articles, the issue of disclosure is related to the sector of deployment of AI. Those authors differentiate between private and public sector applications and demand the disclosure of the algorithmic code in public systems (Bannister and Connolly, 2020: 481; Cerrillo i Martínez, 2019: 18; Zuiderveen Borgesius, 2020: 1584). Bannister and Connolly (2020: 487) even request a prescription of open-source decision-making software for the public sector.

At the same time, disclosure is considered as controversial. Most papers discuss to what extent a reasonable disclosure should be implemented. One consideration is "that there does not exist a credible mechanism by which individuals can readily understand how decisions are being made about them" (Allen and Masters, 2020: 597). Authors share concerns that revealing does not equally mean an understanding of the algorithmic decision at hand (Brand, 2020: 120; Katyal, 2019: 123; Kroll et al., 2017: 638; Rahwan, 2018: 11; Liu et al., 2019: 136). Bloch-Wehba summarizes: "First, disclosure is not only ineffective, it is also legally precluded because these materials are the proper subject of trade secret protections. Second, simply disclosing information about how an algorithm reaches a decision is insufficient to make that information meaningful to the subjects." (Bloch-Wehba, 2020: 1270)

Gacutan and Selvadurai (2020: 201) as well as Cerrillo i Martínez (2019: 21) combine this critique with the demand for a "right to explanation". Others point to the *right to explanation* mentioned in Recital 71 of the GDPR (e.g., Goodman and Flaxman, 2017: 6; Pagallo, 2018: 517; Truby, 2020: 952). Some take it as a role model for legislations that do not guarantee such a right in practice (Hoffmann and Hahn, 2019: 637; Ouchchy et al., 2020: 934; Zerilli et al., 2019: 664; Gacutan and Selvadurai, 2020: 206), others demand a more substantial right to explanation in the EU legislation (Brkan and Bonnet, 2020: 23; Edwards and Veale, 2018: 3; Felzmann et al., 2019: 3).

Furthermore, we see suggestions of regulatory instruments like *labeling* combined with the general *provision of information* for and *education of consumers*. Attached to labeling there is another dimension of disclosure, namely, the disclosure of where AI is used (Carter, 2020: 6; Etzioni, 2018: 31; Hoffmann and Hahn, 2019: 642; Ienca, 2019: 275; Kim, 2020: 932; Redden, 2018: 10).

*4.4.3  Procedural instruments – regulatory actions for the inclusion of actors*

There is a demand to include consumers and affected persons, going beyond information and education. Almost half of the papers point to the need to *encourage the public debate* (e.g., Kroll et al., 2017: 637; Madhavan et al., 2020: 105; Ouchchy et al., 2020: 932; Redden, 2018: 4; Winter and Davidson, 2019: 289; Wong, 2019: 226) and demand the *inclusion of consumers and affected persons* in decision-making processes of where, how, or in which designs AI should be implemented (e.g., Feldstein, 2019: 50; Ienca, 2019: 275; Kaminski, 2019: 1567-1568; Rahwan, 2018: 9; Smith and Desrochers, 2020: 573). These are very broad demands that need further specification. The proposals of how to design inclusion differ substantially: Felzmann et al. (2019: 11) suggest organizing "meeting spaces", and Floridi et al. (2018: 704) consider public consultations to inform policymakers. Shin et al. (2019: 380) and Shneiderman (2020: 6) suggest implementing user experience of system designs.

Besides the regulatory instruments described above in the three subchapters we also identify special and innovative proposals of instruments in the literature. An impressive list of examples is given in Table 1.

Table 1: Innovative proposals for regulatory instruments

| Instrument | Addressed problem(s) | Regulatory Responsibility | Author(s) |
|---|---|---|---|
| Policy Hacking | Overall regulation | Co-Regulation Local affairs and companies | (Abdul Manap & Abdullah 2020, p. 194) (Lauterbach 2019, p. 260) |
| Increase gender diversity in developing teams | Bias and discrimination | Self-regulation (Developers) | (Johnson 2019, p. 1228) (Orr & Davis 2020, p. 732) (Turner Lee 2018, p. 6) |
| Whistleblower protection | Information asymmetry Opacity and accountability | Government | (Katyal 2019, pp. 127-128) |
| Investigative journalism | Information asymmetry Opacity and accountability | Journalists | (Rahwan 2018, p. 11) |
| Financial incentives (surprisingly, as this is a traditional tool of indirect regulation) | Overall regulation | Government | (Kaminski 2019, p. 1575; Floridi et al. 2018, p. 794; Allen & Masters 2020, p. 592) |
| Charter of rights against ADM | Overall regulation | Government | (Bannister & Connolly 2020, p. 487) |

## 4.5  Existing regulations

Although the issues posed by AI entail different possible regulatory instruments, it is important to highlight that the academic debate does not take place in a regulatory vacuum. According to the broad definition of regulation mentioned above, nearly half of the papers review existing provisions related to AI. Firstly, there are legal rules and principles for all human and social activities, including the development of AI. These comprise human rights law, anti-discrimination law, data protection law as well as general values, ethic guidelines and principles. Secondly, regulations with concrete provisions for the protection of affected actors, particularly consumer rights laws, copyright, patent laws, criminal law and tort regulatory systems were mentioned as mandatory rules and principles whose application can be extended to AI through the principle of analogy (Clarke, 2019: 401; Tutt, 2017: 83; Brkan, 2019: 19; Bloch-Wehba, 2020: 1271; Zuiderveen Borgesius, 2020: 1583). In cases where AI software is embedded in devices, product liability is deemed as most important (Clarke, 2019: 402). Thirdly, existing technology-specific laws have been discussed, mainly around industrial robotics or autonomous motor vehicles (Clarke, 2019: 401; Cerrillo i Martínez, 2019: 23; Lauterbach, 2019: 252; Kozuka, 2019: 325; Wang and Siau, 2019: 70; Robles Carrillo, 2020: 7).

Only a few papers identify regulations that are specifically developed for AI, such as particular non-governmentally developed 'Asilomer AI principles' or IEEE standards, ethic guidelines and national strategies (e.g., Clarke, 2019: 401; Floridi et al., 2018: 695; Lauterbach, 2019: 252; Kozuka, 2019: 319-320; Pagallo, 2018: 510; Dignam, 2020: 45). European ethics guidelines „include regulation, standardization, accountability governance, codes of conduct, education, and awareness to foster an ethical mind-set, stakeholder and social dialogue, as well as diversity and inclusive design teams" (Kozuka, 2019: 327).

Ethical AI principles are criticized, because they are "rather vague and fail to give detailed guidance" (Zuiderveen Borgesius, 2020: 1583). Human rights, which are based on more established legal interpretations and practice, at both the universal and national level, should replace ethics as the dominant framework for debate (Robles Carrillo, 2020: 13).

Most authors consider technology-neutral regulations such as the GDPR, that was mentioned in more than half of the papers. This regulation is designed to uphold data protection rights under Article 8 of the EU Charter of Fundamental Rights. It includes a range of measures dealing with the rights of data subjects (individual persons), in particular the right to information, right to access, right not to be subject to automated decisions or to obtain an explanation of decisions reached (e.g., Articles 15, 22 and Recital 71 of the GDPR) (e.g., Brand, 2020: 125; Gasser and Almeida, 2017: 62; Lauterbach, 2019: 251; Zuiderveen Borgesius, 2020: 1579; Kozuka, 2019: 319; Pagallo, 2018: 515; Brkan, 2019: 93; Brkan and Bonnet, 2020: 19; Strandburg, 2019: 1853; Kaminski, 2019: 1551). However, some authors express doubts concerning the effectiveness of the GDPR (Edwards and Veale, 2018: 2; Mann and Matzner, 2019: 3; Goodman and Flaxman, 2017: 5; Treleaven et al., 2019: 34). Brkan focuses on the argument that a right to explanation does not expressly appear in the text of the GDPR, only in the legally non-binding Recital 71 (Brkan, 2019: 115). Zuiderveen Borgesius (2020: 1580) used other words for this inconsistency, and points out that Article 22 of the GDPR is sometimes called the "Kafka provision".

According to some authors, the existing regulations for a disruptive technology such as AI are not well-fitted in principle, because the laws predominantly address specific socio-technical contexts without comprehensive knowledge of new and future challenges (Clarke, 2019: 402; Buiten, 2019: 45; Abdul Manap and Abdullah, 2020: 185; Brand, 2020: 114). Furthermore, existing regulations and the application of laws such as tort and criminal law appear to be inadequate to cope with the substantial threats provoked through AI (Clarke, 2019: 402; Tutt, 2017:

83; Gasser and Almeida, 2017: 61; Nindler, 2019: 34). According to Tutt (2017: 83) only a centralized agency would overcome a piecemeal approach, and would be able to deal with the existing "difficult regulatory puzzles".

Conclusively, many authors argue that existing rules and principles for AI applications need to be revised to better manage the unique characteristics of AI. In particular, data protection (Allen and Masters, 2020: 596; Gurumurthy and Bharthur, 2018: 46; Mann and Matzner, 2019: 3; Bellanova, 2017: 340; Winter and Davidson, 2019: 287), anti-discrimination, human and civil rights (Mann and Matzner, 2019: 2; Cofone, 2019: 1408; Donahoe and Metzger MacDuffee, 2019: 124; Katyal, 2019: 99), and the rules for an explanation (Edwards and Veale, 2018: 4; Brkan and Bonnet, 2020: 23) need to be strengthened and clarified. Some papers propose the introduction of additional new aspects, such as a 'right' to a human decision (Jones, 2017: 219; Johnson, 2019: 1241; Floridi et al., 2018: 697).

Clarke laments that disruptive technologies such as AI genuinely "have features that render existing laws ambiguous and ineffective" and that even "judicial calisthenics" could not solve this problem in a satisfactory manner (Clarke, 2019: 402). Together with others, Clarke requests additional "ex novo" regulatory and normative creations, often expressed as a new coherent and holistic framework (Clarke, 2019: 402; Tutt, 2017: 107-108; Black and Murray, 2019: 16; Floridi et al., 2018: 702-703; Gasser and Almeida, 2017: 61-62; Robles Carrillo, 2020: 15; Wong, 2019: 240; Abdul Manap and Abdullah, 2020: 195).

## 5 Discussion

In the previous chapter we reviewed suggestions for separate aspects of a possible regulation system for AI. In the following analysis we bring together the fragmented pieces of research by providing some overarching observations and evaluations.

First, we observe that the arguments and evaluations of the reviewed papers do not make connections between individual proposed regulatory measures in different application contexts to create a concise and holistic regulation model. The included papers either concentrate on single and heterogenous aspects with contradictory assessments or propose broad meta-models which cannot be implemented easily in practice.

Second, the academic discussion is *centered on industrial states*, with only a few exceptions, e.g., Malaysia (Abdul Manap and Abdullah, 2020). We go along with Smith who laments "that the research is primarily American-based and positions a regulatory approach to algorithms within an American administrative framework" (Smith and Desrochers, 2020: 579), but would add to this view that many authors also refer to EU legislation. Regulatory perspectives from the Global South on the global phenomenon of AI could give valuable insights for an effective regulatory framework.

Third, the academic discourse on regulating AI is a *highly interdisciplinary* research domain. In our final sample, we included articles from law, philosophy, computer science, political science, sociology, and further STEM subjects. Some articles are collective interdisciplinary works, e.g., Floridi et al. include various disciplines in their paper on regulating AI (Floridi et al., 2018). Robles Carrillo stated that only a comprehensive and mutual understanding of the manifold applications, properties and consequences of AI allows decisions about the best possible regu-latory action, which "must be approached from an interdisciplinary perspective" (Robles Carrillo, 2020: 15). Many authors support this position and explicitly demand an increase in (interdisciplinary) research on the societal consequences (e.g., Benkler, 2019: 161; Brkan, 2019: 121; Calo, 2017: 429; Butcher and Beridze, 2019: 88; Cihon et al., 2020: 545; Mannes, 2020: 62). Hence, effective regulation relies on knowledge of the technical specifics as much as on the striking or more subtle societal impacts.

As shown in the analysis of the results we identified recurring topics and salient challenges that come up in the regulatory discussions. We want to reflect some important points by comparing them with the chosen regulatory pathway of the proposed AIA:

First, the academic articles point out to the *core challenge of a shared understanding of AI*, given the different manifestation of applications, as well as the various types of impacts and risks. In Section 4.1 we observe that many authors are aware of the importance of a shared understanding of AI. However, defining AI solely from a technologically based systematization does not frame the regulation object in a sufficient manner, as the choice of a regulatory instrument depends on the dimensions of the risks including the size of the population affected and the severity of harms. In addition, the wide diversity of possible technologies is hard to integrate into a single comprehensive definition or regulation (Robles Carrillo, 2020: 9). In the proposed AIA, the problem of specifying the regulation object was solved by establishing a list of AI technologies and approaches (Annex I) or high-risk AI systems (Annex III) as regulatory objects. According to this approach the regulation of AI will depend on the nature of applications and use cases and its estimated risks. However, further interdisciplinary research could give advice by systematically analyzing AI applications and their risks to fundamental rights and further societal implications. This seems to be a more fruitful approach for advancing the regulatory process, especially to understand the types and dimensions of risks and, thus, to sharpen and extend the scope of regulatory objects, than further engaging in debates about definitions.

Second, there is consensus among the included papers that *existing regulations are not adequate for the substantial and new challenges of AI*, regardless of nation or geographic area, either due to the lack of AI-specific provisions or due to the ineffectiveness of existing regulations, such as data protection law. Here, it must be weighed carefully whether efforts should be invested in improving and tightening existing regulation, or if the drafting of new laws is necessary. This weighing should depend on the perspective of the persons affected and the protection goals of fundamental rights. From this perspective, many deficits of the data protection and anti-discrimination law result from their approach of responsibilities for affected individuals to identify and prove harms and take remedial steps. This approach is no longer feasible in the face of the opacity, adaptability and, partially, technical autonomy of AI systems and the personalization of AI-based services and products. These deficits justify regulatory tasks for agencies as well as holding developers and providers more to account. Both is foreseen in the proposed AIA (see below). However, the GDPR and anti-discrimination legislation can be seen as a catch-all for AI systems that fall through the cracks of the proposed AIA, as it focuses primarily on high-risk applications. However, to fulfil such a task the existent legislation has to amend the included individual rights and be extended by strengthening relevant authorities.

Third, *the academic debate about specific regulatory instruments shows parallels to the European approach* with the proposed AIA. Similarly, both aim to address the fundamental rights and safety risks specific to AI systems through a set of complementary, proportionate, and flexible rules. But we find *significant deviations in the elaboration of the instruments*. The standard instruments identified in the review are pre-market conformity assessment and transparency instruments, which are also included in the European approach. In the literature, restrictive pre-market tools were often considered for specific single applications with defined responsible regulatory actors or bodies. However, there is no consensus whether governmental agencies or market-oriented instruments should control regulatory effectiveness and conformity. This controversy is also salient when looking at the aspect of increased compliance in the case of certification. While some authors generally welcome this instrument, others demand independent oversight and audit for critical control of self-regulation (Kaminski, 2019: 1568). The proposed AIA combines self-assessment and self-certification by providers, voluntary codes-of-conduct with post-market monitoring and controls through providers and MSAs. The European Commission neither took up the scientific argumentation for a single agency (instead made self-regulation by providers a central element for high-risk AI systems), nor even considers it as an option that would have been assessed and compared to other options (European Commission, 2021a: 36-88). It also discarded an option of a regulation of all AI system independent of the risk level, pointing to its estimated high regulatory costs, although the European Commission has stated that this would serve more to protect fundamental rights (European Commission, 2021a: 64-78). Although the MSAs have far-reaching competences, given their wide-ranging tasks their personnel resources envisaged by the European Commission (2021a: Annex 3, 25) seem to be insufficient (Veale and Zuiderveen Borgesius, 2021: 111). Another difference between the scientific literature and the proposed AIA concerns transparency. In the reviewed literature, transparency is most often foreseen as informing affected persons about processes, criteria or impacts of decisions. That information should enable affected persons to take up remedial actions against discriminations or other unjustified treatments. Such a form of disclosure to affected persons is not included in the proposed AIA, nor are individual rights provided for access to adequate information and for correction and redress of unjust treatments. Only AI systems with limited risks require transparency for affected persons in the limited form of information about the existence of the systems, their interaction or about manipulated content (Art. 52 proposed AIA). The publicly accessible EU database, to which providers of high-risk AI systems must register and submit certain information (Art. 60 proposed AIA), will also not contain further information. Additionally, the CE marking (Art. 49 proposed AIA) will not provide such information.

Fourth, in the reviewed literature we often see a missing link between single proposed instruments for specific AI applications and responsibilities and overall regulatory frameworks. Only some authors propose broad hybrid governance frameworks with legal and non-legal dimensions including different regulatory players. They have clear structural ideas of polycentric or layered models, but it remains unclear how to implement different regulatory measures in those metasystems in a robust, holistic, and coherent way. The proposed AIA can also be seen as a hybrid governance framework including monitoring and interventions by governmental agencies, 'regulated self-regulation' approaches in form of standardized self-assessment and self-certification by providers of AI systems, labelling and information instruments, and voluntary codes-of-conduct. It is unclear whether the combination of various elements in the context of the existing legislation can achieve an effective realization of the protection goals. The uncertainty lays within the several objectives of the proposed AIA, in particular, to ensure the internal market and harmonized market rules and at the same time to protect fundamental rights. The coherence of the emerging framework, the interaction of their elements, prevention of regulatory gaps or extensive overlaps and overregulation will remain a continuous task for research.

# 6  Conclusion

This systematic review focuses on the academic debate and the current proposals for the regulation of AI. It provides an overarching regulatory picture by synthesizing the proposals and assessments stemming from literature across different disciplines. Beyond the prevalent assessment that existing legislations are not appropriate for the wide range of AI applications, the results show a partly fragmented, heterogeneous or unspecified field of ideas, and uncovers uncertainty and ambiguities which should serve as a focal point both for further research and political negotiations. Our review did not aim to develop an ideal regulatory model including responsible actors, processes, and instruments. Instead, we have shown that there is already a rich source of scientific knowledge which can be transferred and adopted for regulatory purposes. This stock of knowledge is a viable basis for further research and public discussions that are necessary in many respects, in particular for the evaluation of the emerging regulatory frameworks with respect to effectivity, legitimacy or coherence. Furthermore, the emergence of new AI applications with specific risks requires continuous adaptations of the regulatory frameworks.

In conclusion, we state that the contribution of unfolding the academic debate lies in evolving, strengthening, or criticizing innovative ideas and approaches regarding regulatory instruments, or to support political decisions and the information flow between separate domains. Many questions surrounding the implementation and regulation of AI remain open, as they are fundamental normative decisions of how societies should handle the benefits and harms that result from AI applications. Many authors of the reviewed articles request a widening of narrow discussions, the encouragement of public debate and the inclusion of affected persons in the process for (further) developing regulatory actions. Only by doing justice to the technical as well as the normative and procedural issues we will achieve the often-proclaimed AI for public good.

# Appendix – Search operator

Selection of keywords in the search terms is based on an initial study of the literature and on previous research.

| Data base | Search term | Function |
| --- | --- | --- |
| SCOPUS | ( TITLE-ABS-KEY ( "regulat*" OR "governance" OR "policy mak*" OR "politic*" OR "legislat*") ) | Covering articles about regulation |
|  | TITLE-ABS-KEY ( "AI" OR "artificial intelligen*" OR "algorithm*" OR "ADM" OR "automated decision" OR "machine learning" OR "ML" OR "deep learning" OR "intelligent system" OR "autonomous system" OR "expert system") ) | Covering articles about AI |
|  | AND NOT TITLE-ABS-KEY ( "Bio*" OR "astro*" OR "math*" OR "genetic" OR "physics" OR "sensor*" OR "cancer*" OR "hardware") | Excluding results that are concerned with other issues |
|  | AND ( EXCLUDE ( SUBJAREA , "BIOC" ) OR EXCLUDE ( SUBJAREA , "MATH" ) OR EXCLUDE ( SUBJAREA , "MEDI" ) OR EXCLUDE ( SUBJAREA , "ENER" ) OR EXCLUDE ( SUBJAREA , "AGRI" ) OR EXCLUDE ( SUBJAREA , "PHAR" ) OR EXCLUDE ( SUBJAREA , "MATE" ) OR EXCLUDE ( SUBJAREA , "PHYS" ) OR EXCLUDE ( SUBJAREA , "ENVI" ) OR EXCLUDE ( SUBJAREA , "NEUR" ) OR EXCLUDE ( SUBJAREA , "CHEM" ) OR EXCLUDE ( SUBJAREA , "CENG" ) OR EXCLUDE ( SUBJAREA , "PSYC" ) OR EXCLUDE ( SUBJAREA , "IMMU" ) OR EXCLUDE ( SUBJAREA , "EART" ) OR EXCLUDE ( SUBJAREA , "DENT" ) OR EXCLUDE ( SUBJAREA , "VETE" ) OR EXCLUDE ( SUBJAREA , "HEAL" ) OR EXCLUDE ( SUBJAREA , "NURS" ) OR EXCLUDE ( SUBJAREA , "ENGI" ) OR EXCLUDE ( SUBJAREA , "Undefined" ) ) | Excluding subjects that are concerned with other issues |
|  | AND ( LIMIT-TO ( PUBYEAR , 2021 ) OR LIMIT-TO ( PUBYEAR , 2020 ) OR LIMIT-TO ( PUBYEAR , 2019 ) OR LIMIT-TO ( PUBYEAR , 2018 ) OR LIMIT-TO ( PUBYEAR , 2017 ) OR LIMIT-TO ( PUBYEAR , 2016 ) ) | Limit results to research period January 1st, 2016, to December 31st, 2020 |
|  | AND ( LIMIT-TO ( LANGUAGE , "English" ) ) AND ( LIMIT-TO ( SRCTYPE,"j" ) ) | Limit results to English-language peer-reviewed articles |
| Web of Science | ((TI=( "regulat*" OR "governance" OR "policy mak*" OR "politic*" OR "legislat*" )) OR ( AB=( "regulat*" OR "governance" OR "policy mak*" OR "politic*" OR "legislat*") ) OR ( KP=( "regulat*" OR "governance" OR "policy mak*" OR "politic*" OR "legislat*" ))) | Covering articles about regulation |
|  | AND ((TI=( "AI" OR "artificial intelligen*" OR "algorithm*" OR "ADM" OR "automated decision making" OR "automated decision-making" OR "machine learning" OR "ML" OR "deep learning" OR "intelligent system" OR "autonomous system" OR "expert system" )) OR ( AB=( "AI" OR "artificial intelligen*" OR "algorithm*" OR "ADM" OR "automated decision making" OR "automated decision-making" OR "machine learning" OR "ML" OR "deep learning" OR "intelligent system" OR "autonomous system" OR "expert system" ) ) OR ( KP=( "AI" OR "artificial intelligen*" OR "algorithm*" OR "ADM" OR "automated decision making" OR "automated decision-making" OR "machine learning" OR "ML" OR "deep learning" OR "intelligent system" OR "autonomous system" OR "expert system" ) ) ) | Covering articles about AI |
| | NOT ((TI=( "Bio*" OR "astro*" OR "math*" OR "genetic" OR "physics" OR "sensor*" OR "cancer*" OR "hardware" ) ) OR (AB=( "Bio*" OR "astro*" OR "math*" OR "genetic" OR "physics" OR "sensor*" OR "cancer*" OR "hardware" ) ) OR (KP=( "Bio*" OR "astro*" OR "math*" OR "genetic" OR "physics" OR "sensor*" OR "cancer*" OR "hardware" ))) | Excluding results that are concerned with other issues |
| | AND LANGUAGE: (English) AND DOCUMENT TYPES: (Article) | Limit results to English-language peer-reviewed articles |
| | Refined by: PUBLICATION YEARS: ( 2020 OR 2019 OR 2018 OR 2017 OR 2016 ) | Limit results to research period January 1st, 2016, to December 31st, 2020 |
| | Refined by: WEB OF SCIENCE CATEGORIES: ( COMPUTER SCIENCE ARTIFICIAL INTELLIGENCE OR LAW OR POLITICAL SCIENCE OR COMPUTER SCIENCE SOFTWARE ENGINEERING OR SOCIOLOGY OR SOCIAL SCIENCES INTERDISCIPLINARY OR ROBOTICS OR ETHICS OR SOCIAL ISSUES OR PHILOSOPHY OR PSYCHOLOGY SOCIAL OR SOCIAL SCIENCES BIOMEDICAL ) | Including relevant subjects |
| Worldwide Political Science Abstracts | ti(("regulat*" OR "governance" OR "policy mak*" OR "politic*" OR "legislat*") AND ("AI" OR "artificial intelligen*" OR "algorithm*" OR "ADM" OR "automated decision making" OR "automated decision-making" OR "machine learning" OR "ML" OR "deep learning" OR "intelligent system" OR "autonomous system" OR "expert system")) OR mainsubject(("regulat*" OR "governance" OR "policy mak*" OR "politic*" OR "legislat*") | Covering articles about regulation |
| | AND ("AI" OR "artificial intelligen*" OR "algorithm*" OR "ADM" OR "automated decision making" OR "automated decision-making" OR "machine learning" OR "ML" OR "deep learning" OR "intelligent system" OR "autonomous system" OR "expert system")) OR ab(("regulat*" OR "governance" OR "policy mak*" OR "politic*" OR "legislat*") AND ("AI" OR "artificial intelligen*" OR "algorithm*" OR "ADM" OR "automated decision making" OR "automated decision-making" OR "machine learning" OR "ML" OR "deep learning" OR "intelligent system" OR "autonomous system" OR "expert system")) | Covering articles about AI |
| | *Via filter option* | Excluding results that are concerned with other issues |
| | *Via filter option* | Limit results to English-language peer-reviewed articles |
| | *Via filter option* | Limit results to research period January 1st, 2016, to December 31st, 2020 |
| | *Via filter option* | Including relevant subjects |
| PAIS Index | ti(("regulat*" OR "governance" OR "policy mak*" OR "politic*" OR "legislat*") AND ("AI" OR "artificial intelligen*" OR "algorithm*" OR "ADM" OR "automated decision making" OR "automated decision-making" OR "machine learning" OR "ML" OR "deep learning" OR "intelligent system" OR "autonomous system" OR "expert system")) OR mainsubject(("regulat*" OR "governance" OR "policy mak*" OR "politic*" OR "legislat*") | Covering articles about regulation |

| | AND ("AI" OR "artificial intelligen*" OR "algorithm*" OR "ADM" OR "automated decision making" OR "automated decision-making" OR "machine learning" OR "ML" OR "deep learning" OR "intelligent system" OR "autonomous system" OR "expert system")) OR ab(("regulat*" OR "governance" OR "policy mak*" OR "politic*" OR "legislat*") AND ("AI" OR "artificial intelligen*" OR "algorithm*" OR "ADM" OR "automated decision making" OR "automated decision-making" OR "machine learning" OR "ML" OR "deep learning" OR "intelligent system" OR "autonomous system" OR "expert system")) | Covering articles about AI |
|---|---|---|
| | *Via filter option* | Excluding results that are concerned with other issues |
| | *Via filter option* | Limit results to English-language peer-reviewed articles |
| | *Via filter option* | Limit results to research period January 1st, 2016, to December 31st, 2020 |
| | *Via filter option* | Including relevant subjects |
| Sociological Abstracts | ti(("regulat*" OR "governance" OR "policy mak*" OR "politic*" OR "legislat*") AND ("AI" OR "artificial intelligen*" OR "algorithm*" OR "ADM" OR "automated decision making" OR "automated decision-making" OR "machine learning" OR "ML" OR "deep learning" OR "intelligent system" OR "autonomous system" OR "expert system")) OR mainsubject(("regulat*" OR "governance" OR "policy mak*" OR "politic*" OR "legislat*") AND ("AI" OR "artificial intelligen*" OR "algorithm*" OR "ADM" OR "automated decision making" OR "automated decision-making" OR "machine learning" OR "ML" OR "deep learning" OR "intelligent system" OR "autonomous system" OR "expert system")) OR ab(("regulat*" OR "governance" OR "policy mak*" OR "politic*" OR "legislat*") AND ("AI" OR "artificial intelligen*" OR "algorithm*" OR "ADM" OR "automated decision making" OR "automated decision-making" OR "machine learning" OR "ML" OR "deep learning" OR "intelligent system" OR "autonomous system" OR "expert system")) | Covering articles about regulation and AI |
| | *Via filter option* | Excluding results that are concerned with other issues |
| | *Via filter option* | Limit results to English-language peer-reviewed articles |
| | *Via filter option* | Limit results to research period January 1st, 2016, to December 31st, 2020 |
| | *Via filter option* | Including relevant subjects |